\documentclass[aps,twocolumn,superscriptaddress,preprintnumbers,prl,showpacs]{revtex4-1}

\usepackage{graphicx}
\usepackage{dcolumn}
\usepackage{bm}
\usepackage{epsfig}
\usepackage{color}

\usepackage{longtable}
\usepackage{amsmath}

\begin{document}

\preprint{Submitted to {\it Physical Review Letters}}

\title{Macro deformation twins in single-crystal aluminum}

\author{F.~Zhao}
\affiliation{The Peac Institute of Multiscale Sciences, Chengdu, Sichuan 610031, P. R. China}

\affiliation{Key Laboratory of Advanced Technologies of Materials, Ministry of Education, Southwest Jiaotong University, Chengdu, Sichuan 610031, P. R. China}

\author{L.~Wang}
\affiliation{The Peac Institute of Multiscale Sciences, Chengdu, Sichuan 610031, P. R. China}

\author{D.~Fan}
\affiliation{The Peac Institute of Multiscale Sciences, Chengdu, Sichuan 610031, P. R. China}

\author{B. X.~Bie}
\affiliation{The Peac Institute of Multiscale Sciences, Chengdu, Sichuan 610031, P. R. China}
\affiliation{School of Science, Wuhan Univiersity of Technology, Wuhan, Hubei 430070, P. R. China}

\author{X. M.~Zhou}
\affiliation{The Peac Institute of Multiscale Sciences, Chengdu, Sichuan 610031, P. R. China}

\author{T.~Suo}
\affiliation{School of Aeronautics, Northwestern Polytechnical University, Xi'an, Shaanxi 710072, P. R. China}

\author{Y. L.~Li}
\affiliation{School of Aeronautics, Northwestern Polytechnical University, Xi'an, Shaanxi 710072, P. R. China}

\author{M. W.~Chen}
\affiliation{WPI Advanced Institute for Materials Research, Tohoku University, Sendai 980-8577, Japan}
\affiliation{State Key Laboratory of Metal Matrix Composites, School of Materials Science and Engineering, Shanghai Jiao Tong University, Shanghai 200030, P. R. China}

\author{C.~Liu}
\affiliation{Institute of Fluid Physics, Mianyang, Sichuan 621900, P. R. China}

\author{M. L.~Qi}
\affiliation{School of Science, Wuhan Univiersity of Technology, Wuhan, Hubei 430070, P. R. China}

\author{M. H.~Zhu}
\email{zhuminhao@swjtu.cn}
\affiliation{Key Laboratory of Advanced Technologies of Materials, Ministry of Education, Southwest Jiaotong University, Chengdu, Sichuan 610031, P. R. China}

\author{S. N.~Luo}
\email{sluo@pims.ac.cn}
\affiliation{The Peac Institute of Multiscale Sciences, Chengdu, Sichuan 610031, P. R. China}

\date{\today}

\begin{abstract}
Deformation twinning in pure aluminum has been considered to be a unique property of nanostructured aluminum. A lingering mystery is whether deformation twinning occurs in coarse-grained or single-crystal aluminum, at scales beyond nanotwins. Here, we present the first experimental demonstration of macro deformation twins in single-crystal aluminum formed under ultrahigh strain-rate ($\sim$10$^6$ s$^{-1}$), large shear strain (200$\%$) via dynamic equal channel angular pressing. Deformation twinning is rooted in the rate dependences of dislocation motion and twinning, which are coupled, complementary processes during severe plastic deformation under ultrahigh strain rates.
\end{abstract}

\maketitle

When we talk about crystal deformation, what do we actually talk about? Crystal defects \cite{greer13nm}. Crystal defects such as dislocations (line defects) and twins (planar defects) play a critical role in plastic deformation and ultimately govern the multifarious mechanical behaviors of many crystalline materials \cite{christian95pms, zhu12pms}. While both dislocation slip and deformation twinning are dependent on an intrinsic material property -- stacking fault energy \cite{swygenhoven04nm, sarma10msea} (SFE), their sensitivities to SFE differ considerably. A notable example is pure aluminum, a typical face-centered cubic (fcc) metal with high SFE (104--142 mJ\,m$^{-2}$) \cite{yamakov02nm}, in which deformation twinning rarely occurs even deformed at low temperatures and/or at high strain rates \cite{jiang09sm, liu98am}. This rareness of deformation twinning in such materials is normally attributed to the following two reasons: (i) a large number of slip systems in fcc metals render dislocation slip a very efficient deformation mode \cite{daphalapurka14am, blewitt57jap}, and (ii) the nucleation of twinning partial dislocations require much higher shear stresses than trailing partial dislocations due to the high unstable twin fault energy \cite{warner07nm}. Searching for macro deformation twins in pure aluminum and revealing the underlying mechanisms have been of sustained interest in the past decade.

Molecular dynamics (MD) simulations first predicted that nanoscale deformation twins can nucleate under high tensile stress (2.5 GPa) and high strain rate (10$^7$ s$^{-1}$) in nanograined aluminum \cite{yamakov02am, yamakov02nm}, and subsequent experiments confirmed this prediction in nanograined aluminum films under different kinds of severe plastic deformation (SPD) \cite{chen03sci, liao03apl, liao03apl1}. One explanation was proposed based on classical dislocation theory \cite{hirth92b}: when grain size decreases to tens of nanometers, normal dislocation activities are greatly suppressed by the high fraction of grain boundaries (GBs); as a result, deformation twinning takes over as the dominant deformation mechanism \cite{chen03sci}. Besides nanograin size effect, many simulation and experimental studies suggest that deformation twinning prefers to occur at high strain rates in fcc metals \cite{meyers01am, xiao08sm}. This rate-dependent twinning mechanism has been corrobarated by a very recent experiment on pure aluminum with comparatively large nanograins (50-100 nm) \cite{cao15meccanica}. However, there has been no solid evidence for deformation twinning in single-crystal or coarse-grained pure aluminum. It is natural to ask whether deformation twinning indeed occurs  in coarse-grained and single-crystal aluminum, and if it does, whether the mechanisms are the same as those in nanostructured aluminum. In other words, is deformation twinning just unique to nanostructured aluminum?

In the present work, we develop a novel dynamic equal channel angular pressing (D-ECAP) technique, schematically illustrated in Fig.~1a (supplementary materials Fig.~S1),  to enable large shear plastic deformation at an ultrahigh strain rate to investigate twinning in single-crystal aluminum. In traditional ECAP method, the piston is pressed at a low speed (10$^{-3}$ m\,s$^{-1}$), while in D-ECAP process, a much higher piston velocity (10$^2$ m\,s$^{-1}$) is applied to induce ultrahigh strain rates (10$^6$ s$^{-1}$).

\begin{figure}[t]
\centering
\includegraphics[scale=0.9]{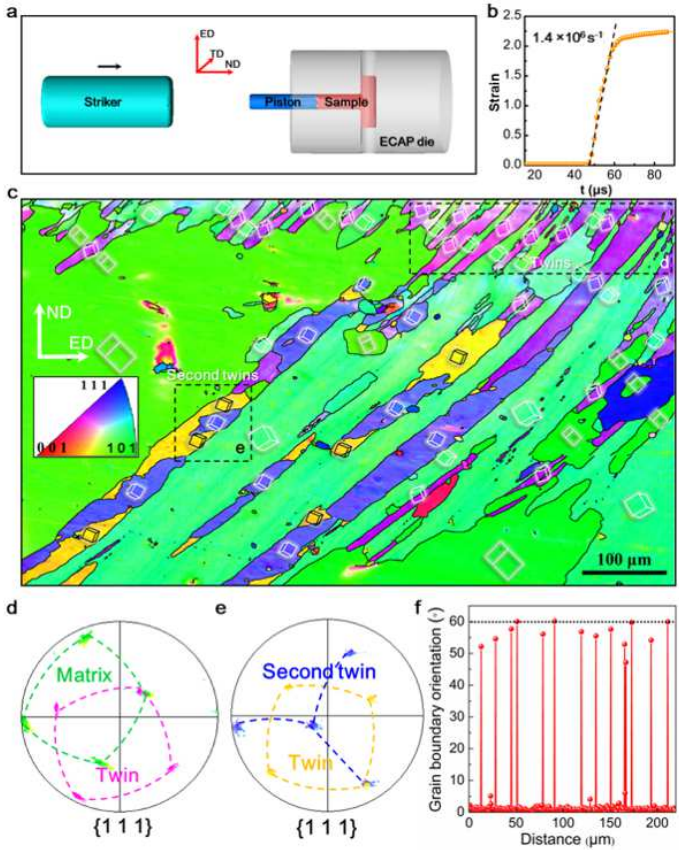}
\caption{Dynamic equal channel angular pressing (D-ECAP) experiments, and microstructure overview of macro deformation twins in single-crystal aluminum. {\bf a}, Schematic of the D-ECAP setup. {\bf b}, Evolution of plastic strain in the specimen under D-ECAP, and the estimated strain rate (dashed line). {\bf c}, Typical inverse pole figure (IPF) map of high-density macro deformation twins. ({\bf d} and {\bf e}) Pole figures of deformation twins and secondary twins corresponding to regions {\bf d} and {\bf e} marked in {\bf c}. {\bf f}, Twin boundary misorientation profile in region {\bf d}. }
\end{figure}

\begin{figure}[t]
\centering
\includegraphics[scale=0.5]{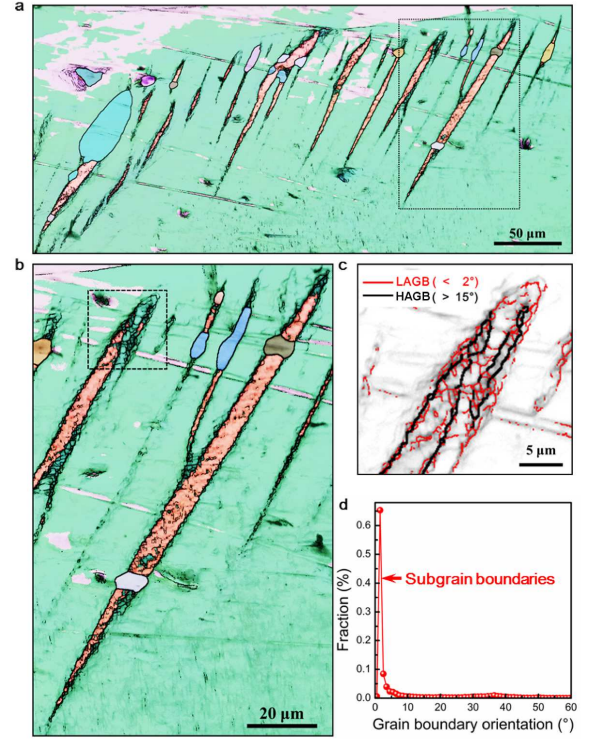}
\caption{Deformation twin boundaries, low-angle grain boundaries (LAGBs), and high-angle grain boundaries (HAGBs) in an aluminum specimen after D-ECAP. {\bf{a}}, EBSD image of deformation twins formed in the interior of the specimen. {\bf b}, Distribution of LAGBs (black lines) around deformation twins. {\bf c}, Magnified image of the region delimited by the rectangle in {\bf b} highlighting interactions between twin boundaries and LAGBs. Twin boundaries can act as dislocation sources and barriers to dislocation motion. {\bf d}, Misorientation distribution corresponding to {\bf b}, indicating that LAGBs or high-density dislocation structures are the major deformation microstructure within deformation twins. }
\end{figure}

A single-crystal aluminum rod (ED--[010], TD--[$\bar{1}$01], ND--[101], Fig.~1a, c) with a purity of 99.999\%, 3 mm in diameter, is set in the D-ECAP die, and then extruded for only one pass by high velocity pressing at room temperature. The estimated maximum strain rate from finite elemental analysis is about 1.4$\times$10$^6$ s$^{-1}$ (Fig.~S2) and the corresponding plastic strain is about 200$\%$ (Fig.~1b). After extrusion via D-ECAP, the specimen is characterized with electron back-scattering diffraction (EBSD). High-density macro deformation twins are observed near the upper extrusion surface (Fig.~1c, S3, S4). Such macro deformation twins had never been found previously in pure aluminum, either in nanocrystalline, ultrafine, coarse-grained, or single-crystal aluminum.

An example of EBSD analysis of the recovered specimen after D-ECAP is shown in an inverse pole figure or IPF map (Fig.~1c). The average twin width is about 40 $\mu$m and the maximum length is about 500 $\mu$m. The twin boundaries are determined to be typical $\Sigma$3\{111\}60$^{\circ}$ coherent twin boundaries (Fig.~S3). Figure~1c also shows that all the macro deformation twins nucleate from the upper surface where shear deformation is augmented by the sample--die friction, and propagate toward sample interior along the shear direction ([$\bar{1}$21]). However, unlike normal straight twin boundaries, the deformation twin boundaries induced by D-ECAP process are curved, indicating twin boundary rotation. This is caused by strong interactions between twins and dislocations under severe shear plastic deformation. A large fractions of low angle grain boundaries (LAGBs), which correspond to high-density dislocation structures, are found in the vicinity of twin boundaries (Figs.~S5, S6). Occasionally, secondary deformation twins form at preceding twin boundaries (Fig.~1c). These primary and secondary deformation twins are identified from pole figure maps, and the twin planes are \{111\} (Figs.~1d, 1e). Grain boundary orientation analysis reveals that the average misorientations are approximately 60$^{\circ}$, and the slight deviations from this angle are caused by twin boundary rotation (Fig.~1f).

The deformation twins in single-crystal aluminum processed with D-ECAP are ``macroscopic'' in nature, distinct from nanotwins observed previously in nanograined aluminum \cite{chen03sci, liao03apl, liao03apl1}. In nanograined aluminum, as grain size decreases to below a critical value ($\sim$15 nm), the critical stress for dislocation nucleation becomes higher than that for deformation twinning, and therefore, deformation twinning replaces dislocation slip as the preferred deformation mode \cite{chen03sci}. However, nanoscale grain boundaries can also strongly restrict twin propagation, thus limiting twin size to several nanometers. Our experiments demonstrate that deformation twins can be induced by simultaneous ultrahigh strain rates (via dynamic loading) and large shear deformation (via ECAP), and grow to sub-mm scales in single-crystal aluminum without grain boundaries serving as the barrier to twin propagation. MD simulations suggest that spontaneous self-pinning of dislocations impedes the motion of dislocations at high strain rates, and thus deformation twinning takes over as the predominant deformation mode. Our experimental results confirm this prediction \cite{marian04nm}.

Figure~2a shows abundant deformation twins with a typical lens shape formed within the sample interior, which has undergone less friction and smaller shear strain. During severe plastic deformation, high-density dislocation structures (LAGBs) are found to cluster around twin boundaries, suggesting that deformation twins can serve as dislocation sources as well as barriers, effectively hindering dislocation motion around them (Figs.~2b, 2c). To further verify this observation, twin and grain boundary fractions are quantified from the IPF map. As shown in Fig.~2d, LAGBs, or high-density dislocation structures, are the most prominent microstructure feature. During such ultrahigh strain-rate shear deformation, dislocation nucleation and interactions are still significant near and within deformation twin boundaries, and the propagation of deformation twins assists dislocation nucleation as new dislocation sources (Fig.~S5). Our experiments provide direct evidence that dislocation slip and deformation twinning can act together as complementary deformation modes during ultrahigh strain-rate shear deformation. This is contrary to previous MD simulations and experiments claiming that the transition from dislocation slip to twinning can occur in aluminum only when dislocation activities are strongly suppressed.

In order to help reveal phenomena and underlying mechanisms related to deformation twinning under ultrahigh strain-rate shear deformation within the context of D-ECAP experiments, MD simulations are carried out. The simulation details are presented in supplementary materials. In our MD simulations mimicking the D-ECAP experiments, the pressing velocity is relatively low (10 m\,s$^{-1}$), but the corresponding strain rate is still high ($\sim$10$^8$ s$^{-1}$) given the relatively small length scales. When the sample initially passes through the die corner, the leading partial dislocations first emit from this area (Fig.~3a). The resolved shear stress is about 1 GPa (1.05$\%$, Fig.~S8), comparable to twin nucleation stress reported previously, 0.64 GPa \cite{chen03sci} and 0.93 GPa \cite{yamakov05am, koning03prl}. Consequently, the twin partial dislocations immediately appear following the leading partial dislocations, and then a nanotwin forms. Figure 3b and 3c show that, as deformation twins propagate, dislocations nucleate from the twin boundaries, and this greatly enhances the local deformation ability in order to accommodate the high deformation rate. On the other hand, dislocations quickly become self-pinned and tangled (Fig.~S9) as they move away from twin planes. High-density dislocation structures (i.e., LAGBs) arise near the twin boundaries, consistent with our experimental results. When the propagation of deformation twins ceases, dislocation motion and dislocation interactions begin to dominate plastic deformation, and annihilation of certain twins occurs as a result (Fig.~3d).

\begin{figure}[t]
\centering
\includegraphics[scale=0.37]{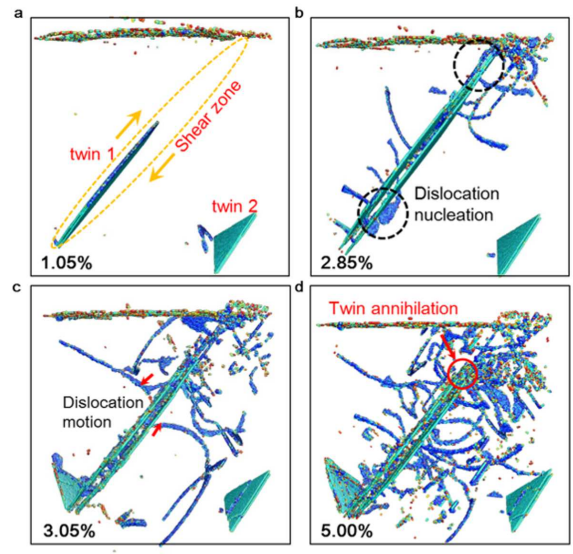}
\caption{Snapshots of deformation twinning and interactions between twin boundaries and dislocations, obtained from MD simulations. A movie of the simulations is included in the supplementary materials. {\bf{a}}, Deformation twin nucleation and propagation in the shear zone. {\bf b}, Dislocation nucleation from twin boundaries, as indicated by circles. {\bf c}, Dislocation motion around twin boundaries. {\bf{d}}, Twin annihilation caused by the interactions between twin boundaries and dislocations. }
\end{figure}

\begin{figure}[t]
\centering
\includegraphics[scale=0.32]{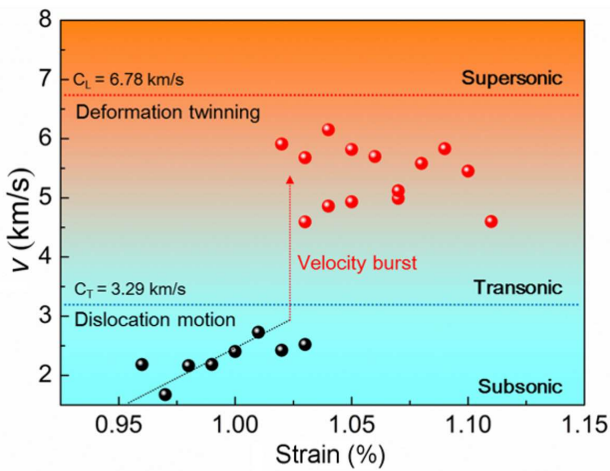}
\caption{Propagation speed of dislocations and deformation twins during D-ECAP obtained from MD simulations. }
\end{figure}

It has been shown that deformation twins propagate faster at high loading rate in a number of materials \cite{daphalapurka14am, faran10prl, abeyarantne03jmps}. Our D-ECAP experiments and MD simulations demonstrate the coexistence of deformation twinning and dislocation slip in single-crystal aluminum. Therefore, it is desirable to compare the propagation speeds of dislocation motion and deformation twinning during high-rate shear deformation. Figure~4 shows that the leading partial dislocations nucleate first, and the speed of dislocation motion increases with increasing shear strain but remains subsonic. Subsequent transition to deformation twinning leads to a velocity burst, and deformation twins propagate at a transonic speed. Classical dislocation theory predicts that, dislocation velocity cannot exceed the shear wave velocity \cite{hirth92b}, and with increasing dislocation velocity, dislocation self-pinning and tangling ensue, leading to stress accumulation until deformation twins nucleate \cite{bulatov06nat, devincre08sci}. Twin propagation velocity exceeds the shear wave velocity. The enormous propagation velocity jump (subsonic to transonic) is a direct result of the dislocation-twinning transition in single-crystal aluminum necessitated by the ultra-high strain rate.

The D-ECAP experiments and MD simulations reveal that deformation twinning and dislocation slip coexist in single-crystal aluminum under ultrahigh strain-rate loading. The applied strain rate, $\dot{\varepsilon}_{\rm applied}$, consists of contributions from dislocation motion and deformation twinning:
\begin{equation}
 \dot{\varepsilon}_{\rm applied} = \dot{\varepsilon}_{\rm dislocation} + \dot{\varepsilon}_{\rm twin} .
\end{equation}
Based on classical dislocation theory \cite{hirth92b}, $\dot{\varepsilon}_{\rm dislocation}$ = $mb \rho v$, where $m$ is a material-specific parameter, $b$ is the Burgers vector amplitude, $\rho$ is the mobile dislocation density, and $v$ is the average dislocation propagation velocity. Twin boundaries can serve as new dislocation sources while propagating, and subsequently, the mobile dislocation density increases rapidly due to fast twin propagation. Then Eq. (1) can be rewritten as:
\begin{equation}
  \dot{\varepsilon}_{\rm applied} = mbv(\rho_0+NLv_{\rm twin}t^2) + \dot{\varepsilon}_{\rm twin},
\end{equation}
where $\rho_0$ is the initial mobile dislocation density, $N$ is dislocation nucleation rate, $v_{\rm twin}$ is the propagation velocity of a deformation twin, and $L$ is the average width of twins (Fig.~S10, supplementary materials).

If deformation involves dislocation motion only, then $\dot{\varepsilon}_{\rm dislocation}=mbv\rho_0$, and is limited by $v$ which is subsonic. As a result, dislocation itself cannot accommodate the applied strain rate for sufficiently high $\dot{\varepsilon}_{\rm applied}$, and the strain deficit has to be supplemented by faster propagating deformation twinning (transonic twinning, Fig.~4). Dislocation self-pinning and tangling slow down dislocation motion, and give rise to intense stress accumulation and subsequently deformation twinning. New dislocations nucleate at propagating twin boundaries (as new dislocation sources) \cite{zhu08prl, fan12prl},  and mobile dislocation density increases by $NLv_{\rm twin}t^2$. Thus, dislocation frustration leads to deformation twinning, which in turn assists dislocation nucleation. In other words, dislocation slip and deformation twinning coexist and complement each other necessarily in high-rate plastic deformation of aluminum, as seen in our experiments and MD simulations.

Our D-ECAP experiments and MD simulations have demonstrated that deformation twinning is not unique to nanostructured aluminum, and macro deformation twinning can occur in single-crystal, and likely coarse-grained, aluminum under ultrahigh strain rates.  Deformation twinning and dislocation slip are rate-controlled, coupled, and complementary processes during ultrahigh strain-rate plastic deformation. Deformation twinning, once elusive in experiments, may indeed play a major role in plastic deformation under high strain rates for aluminum, and conceivably, other high-SFE metals.

\section*{Acknowledgement}
The work was partially supported by the 973 Project (Grant No. 2014CB845904), and NSF and NSAF of China (Nos. 11472253 and U1230202).



\end{document}